\documentclass{article}
\usepackage{ijcai21}

\usepackage{times}
\usepackage{soul}
\usepackage{url}
\usepackage[hidelinks]{hyperref}
\usepackage[utf8]{inputenc}
\usepackage[small]{caption}
\usepackage{graphicx}
\usepackage{amsmath}
\usepackage{amsthm}
\usepackage{booktabs}
\usepackage{algorithm}
\usepackage{algorithmic}
\usepackage{subfigure}
\usepackage{threeparttable}
\usepackage{multirow}
\usepackage{multicol}
\usepackage{bbding}
\usepackage{tikz}
\usepackage{calc}
\usepackage{amssymb}
\usepackage{caption}
\usepackage{colortbl}

\usepackage[group-separator={,}]{siunitx}
\title{Dialogue Disentanglement in Software Engineering: How Far are We?}

\author{
Ziyou Jiang$^{1,3}$
\and
Lin Shi$^{1,3,}$\thanks{Corresponding author.}
\and
Celia Chen$^4$
\and
Jun Hu$^{1,3}$
\And
Qing Wang$^{1,2,3}$
\affiliations
$^1$Laboratory for Internet Software Technologies, $^2$State Key Laboratory of Computer Sciences,\\
Institute of Software Chinese Academy of Sciences, Beijing, China\\
$^3$University of Chinese Academy of Sciences, Beijing, China\\
$^4$Department of Computer Science, Occidental College, Los Angeles, California, USA\\
\emails
\{ziyou2019, shilin, hujun, wq\}@iscas.ac.cn,
qchen2@oxy.edu
}

\begin{document}

\maketitle

\begin{abstract}

Despite the valuable information contained in software chat messages, disentangling them into distinct conversations is an essential prerequisite for any in-depth analyses that utilize this information. To provide a better understanding of the current state-of-the-art, we evaluate five popular dialog disentanglement approaches on software-related chat. We find that existing approaches do not perform well on disentangling software-related dialogs that discuss technical and complex topics. Further investigation on how well the existing disentanglement measures reflect human satisfaction shows that existing measures cannot correctly indicate human satisfaction on disentanglement results. Therefore, in this paper, we introduce and evaluate a novel measure, named \textit{DLD}. Using results of human satisfaction, we further summarize four most frequently appeared bad disentanglement cases on software-related chat to insight future improvements. These cases include (i) ignoring interaction patterns; (ii) ignoring contextual information; (iii) mixing up topics; and (iv) ignoring user relationships. We believe that our findings provide valuable insights on the effectiveness of existing dialog disentanglement approaches and these findings would promote a better application of dialog disentanglement in software engineering.

\end{abstract}

\section{Introduction}
Online communication platforms such as Gitter\footnote{\url{www.gitter.im}} have been heavily used by software organizations as the collaborative team communication tools for their software development teams. 
Conversations on these platforms often contain a large amount of valuable information that may be used, for example, to improve development practice, to observe software development process, and to enhance software maintenance. However, these conversations oftentimes appear to be entangled. While some developers may participate in active discussions on one topic, others could address previous conversations on different topics in the same place. Without any indications of separate conversations, it is difficult to extract useful information when all messages appear together. 

A number of approaches has been proposed to address such issue of dialog entanglement. Early work uses simple classifiers with handcrafted features to analyze the coherence of message-pairs~\cite{disentangle1}. In addition, neural network is used to obtain the relationships with the development of deep learning, such as \textit{FeedForward} (\textit{FF})~\cite{acl19disentangle}, \textit{CNN}~\cite{jiangdisentangle} and \textit{BiLSTM}~\cite{bilstm2017usage} {\it etc.}. Meanwhile, sequential encoder/decoder based models are recently introduced with pre-trained \textit{BERT}~\cite{li2020dialbert}, session-state dialog encoders~\cite{ijcai20e2e}, and pointer networks~\cite{emnlp20disentangle}. 
To evaluate their approaches, most of these studies use conversations mined from social platforms, such as Reddit and Twitter. The content of these conversations usually focuses on general topics, such as movies and news. 


{Unlike general conversations, software-related conversations have different and distinct characteristics: (1) SE dialogs heavily focus on resolving issues, thus they are mostly in the form of question and answer. In our observation, nearly 90\% of the SE dialogs are in this Q\&A style. However, in the general dialogs (i.e. the  news or movie dialogs), the proportion of such Q\&A dialogs is much less. Existing approaches that do not consider this Q\&A aspect may result in disentanglement errors (elaborated in Section 5). (2) SE dialogs are domain-specific and each domain has its own technical terms and concepts. Existing approaches that leverage general dialogs without incorporating any domain-specific terminology cannot understand jargons accurately. (3) SE dialogs usually involve more complex problems, which require developers to discuss various topics within one dialog. Thus, SE dialogs are more likely to show a higher degree of entanglement.}

However, the disentanglement performance on software-related chat has rarely been evaluated. ~\cite{acl19disentangle} and ~\cite{emnlp20disentangle} evaluate their models on a set of Ubuntu IRC dataset but the generalizability of the results is limited due to the small sample size. Moreover, although the goal of dialog disentanglement is to provide users with the ease of finding valuable information from entangled messages, none of the previous studies investigate how well existing disentanglement measures reflect human satisfaction. Thus, it remains unclear how far we are from effective dialog disentanglement towards software-related chat that contains the majority of technical and professional conversations.

In this paper, we conduct an exploratory study on 7,226 real-world developers' dialogs mined from eight popular open-source projects hosted on Gitter. 
First, we compare five state-of-the-art dialog disentanglement approaches based on two strategies: transferring the original models across domains and retraining the models on software-related dialogs. Second, we further investigate how well the existing disentanglement measures reflect human satisfaction. The results show that the existing measures are unable to accurately indicate human satisfaction on the disentanglement results. Therefore, we propose and evaluate a novel measure that measures human satisfaction on disentanglement results by incorporating Levenshtein distance, named \textit{DLD}, to complement the existing literature. To further understand why existing approaches fail to disentangle software-related chat, we summarize four bad cases of the incorrect disentanglement. We believe that the findings we have uncovered will promote a better application of dialog disentanglement in the software engineering domain.
The major contributions of this paper are as follows:

$\bullet$ We conduct a comparative empirical study on evaluating the state-of-the-art disentanglement approaches on software-related chat;

$\bullet$ We propose a novel measure, \textit{DLD}, for {quantitatively} measuring human satisfaction on disentanglement results;

$\bullet$ We release a dataset\footnote{\url{https://github.com/disensoftware/disentanglement-for-software}} of disentangled software-related dialogs to facilitate the replication of our study and future improvements of disentanglement models.







\section{Dataset Preparation}

\textbf{Study Projects.}
Gitter is selected due to its high popularity and being openly accessible. A sample dataset is constructed from the most participated projects found in eight popular domains, including front-end framework, mobile, data science, DevOps, blockchain platform, collaboration, web app, and programming language. The total number of participants across these eight projects is 95,416, which accounts for 13\% of the entire Gitter's participant population.



\noindent\textbf{Data Preprocessing.}
For each project, we collect all the {1,402,894} utterances recorded before ``2020-11-20''. Data is then preprocessed by the following steps: 

(1) \textit{Formatting}. To preprocess the textual utterances, we first normalize the non-ASCII characters (\textit{i.e.,} emojis) to standard ASCII strings. Since low-frequency tokens such as URL, email address, code, HTML tags, and version numbers do not contribute to the classification results in chat utterances, we replace them with specific tokens \textit{[URL], [EMAIL], [HTML], [CODE]} and \textit{[ID]} respectively. We also utilize Spacy\footnote{\url{https://www.spacy.io/}} to tokenize sentences into terms. To alleviate the influence of word morphology, we then perform lemmatization and lowercasing on terms with Spacy.
{(2) \textit{Experiment Dataset Creation}. 
We employed a 3-step manual process to generate the dialog dataset for our experiments.
First, we randomly sampled 100 utterances from each community's live chat log, intending to trace corresponding dialogs associated with each utterance. This step led to a total of 800 utterances sampled from eight projects. Next, using each utterance as a seed, we identified its preceding and succeeding utterances iteratively and grouped the related utterances into the same dialog. We used these manually disentangled dialogs as the \textbf{Ground-Truth} Data. Finally, we added the intervened utterances that posted among those labeled dialogs back into the dataset.}
(3) \textit{Exclusion}. We exclude unreadable dialogs, including a) dialogs that are written in non-English languages; b) dialogs that contain too many specific tokens; and c) dialogs with typos and grammatical errors. 
The final dataset contains 749 disentangled dialogs consisting of 7,226 utterances, contributed by 822 participants. Detailed statistics are shown in Table \ref{project_prepration}.

\begin{table}[t]
\centering
 \resizebox{0.9\columnwidth}{!}{
\begin{tabular}{c|c|c|c|c|c|c|c}
\toprule
\multirow{2}{*}{Id}&\multirow{2}{*}{Project} & \multirow{2}{*}{Domain} & \multicolumn{2}{c|}{Entire Population} & \multicolumn{3}{c}{Sample Population} \\
\cline{4-8}
& & &PA & UT & PA & DL & UT \\ \hline
P1&Angular          & Frontend Framework         & 22,467                       & 695,183              & 125                       & 97                       & 778                         \\ \hline
P2&Appium          & Mobile                     & 3,979                          & 29,039               & 73                                      & 87                       & 724                         \\ \hline
P3&Dl4j            & Data Science               & 8,310                          & 252,846              & 93                       & 100                       & 1,130                         \\ \hline
P4&Docker           & DevOps                     & 8,810                        & 22,367               & 74               & 90                       & 1,126                         \\ \hline
P5&Ethereum       & Blockchain Platform & 16,154                       & 91,028               & 116                       & 96                       & 516                         \\ \hline
P6&Gitter         & Collabration Platform      & 9,260                           & 34,147               & 87                       & 86                       & 515                         \\ \hline
P7&Typescript       & Programming Language                   & 8,318              & 196,513              & 110              & 95                       & 1,700                         \\ \hline
P8&Nodejs          & Web-App Framework  & 18,118                          & 81,771               & 144                       & 98                       & 737                         \\ \hline
\multicolumn{3}{c|}{\textit{Total}}          & \textit{95,416}      & \textit{1,402,894}    & \textit{822}             & \textit{749}             & \textit{7,226}               \\ \bottomrule
\end{tabular}
}
\caption{The characteristics of selected projects. 
{\it PA} represents the number of participants,
{\it DL} represent the number of dialogs, {\it UT} represents the number of utterances.}
\label{project_prepration}
\end{table}



\section{Model Comparison}
In this section, we conduct an in-depth empirical study to evaluate the state-of-the-art dialog disentanglement models towards software-related chat.


\subsection{Model Selection}
By searching through the literature published in the representative venues (\textit{Computational Linguistics, NAACL, ACL, IJCAI, EMNLP and SIGIR}) for the last 15 years, we choose eight approaches as the candidate models for our study. Their code accessibility, dataset accessibility, and learning technologies are summarized in Table \ref{models_metrics_selection}. From the table, we can observe that five out of the eight models provide public assess to their source code and dataset. Moreover, these five models also utilize more advanced technologies such as deep neural network. Thus, we select {\it FF}, {\it BiLSTM}, {\it BERT}, {\it E2E} and {\it PtrNet} as the state-of-the-art (SOTA) models to be compared in our study. Each model is described in detail as follows.

Given a graph of utterances $G\{V,E\}$, the goal of \textit{FF model} and \textit{BiLSTM model} is to predict whether there is an edge $E$ between the utterance $u_j$ and $u_k$, where utterances are nodes and edges
indicate that one utterance is a response to another. Each connected component is a conversation.   
\begin{equation}
 G=\{V,E\}, V=\{u_1,u_2,...u_n\}, E=<u_j,u_k>
\end{equation}

(1) \textbf{FF model} is a feedforward neural network with two layers, 256 dimensional hidden vectors, and softsign non-linearities. The input is a 77 dimensional numerical feature extracted from the utterance texts, which includes TF-IDF, user name, time interval, whether two utterances contain the same words and \textit{etc.}. \textit{FF} model is trained from 77,563 manually annotated utterances.

(2) \textbf{BiLSTM model} is a bidirectional recurrent neural network with 160 context maximum size, 200 neurons with one hidden layer. The input is a sequence of 512 dimensional word vectors. \textit{BiLSTM} model is trained from 7.1 million utterances and 930K dialogs.

Unlike \textit{FF} and \textit{BiLSTM}, given a sequence of utterances $[u_1, u_2, ... , u_n]$, the goal of \textit{BERT}, \textit{E2E} and \textit{PtrNet} is to learn the following probability distribution, where $y_i$ is the action decision for the utterance $u_i$.
\begin{equation}
    P(D) = \prod_{i=1}^{n}P(y_i|y_{<i}, u_{\leq i}); 1\leq i \leq n
\end{equation}

(3) \textbf{BERT model} uses the \textit{Masked Language Model} and \textit{Next Sentence Prediction}~\cite{bertbase} to encode the input utterances, with 512 embedding size and 256 hidden units. \textit{BERT} model is trained from 74,963 manually annotated utterances.

(4) \textbf{E2E model} performs the dialog Session-State encoder to predict dialog clusters, with 512 embedding size, 256 hidden neurons and 0.05 noise ratio. \textit{E2E} model is trained from 56,562 manually annotated utterances.

(5) {\textbf{PtrNet model} utilizes the pointer network to predict links within utterances, with 512 embedding size and 256 hidden units. \textit{PtrNet} model is trained from 52,641 manually annotated utterances.}

\begin{table}[t]
\centering
\begin{threeparttable}
 \resizebox{0.9\columnwidth}{!}{
\begin{tabular}{p{4.6cm}<{\centering}|c|c|p{3.8cm}<{\centering}}
\toprule
Model & Code Available & Dataset Available & Technology \\
\hline
Weighted-SP~\cite{shen-F} & No & No (Linux) & Weight Calculation  \\
\hline
ME Classifier~\cite{disentangle1}  & No & No (Ubuntu) & Traditional Classifier\\
\hline 
BiLSTM~\cite{bilstm2017usage} & Yes & Yes (Movie) & Recurrent Neural Network \\
\hline 
CISIR~\cite{jiangdisentangle} & No & Yes (News) & Convolutional Neural Network \\
\hline 
FF~\cite{acl19disentangle} & Yes & Yes (Ubuntu) & FeedForward Neural Network\\
\hline 
BERT~\cite{li2020dialbert}  & Yes & Yes (Movie) &  Encoder/Decoder Network\\
\hline 
E2E~\cite{ijcai20e2e} & Yes & Yes (Movie) & Encoder/Decoder Network\\
\hline 
PtrNet~\cite{emnlp20disentangle} & Yes & Yes (Ubuntu) & Encoder/Decoder Network\\
\bottomrule
\end{tabular}}
\end{threeparttable}
\caption{Candidate disentanglement models.}
\label{models_metrics_selection}
\end{table}

\subsection{Evaluation Measures}

We investigate the evaluation measures that are adopted by existing literature, as shown in Table \ref{metrics_selection}. There are four evaluation measures that are frequently used: 
{\it ARI}~\cite{SantosE09}, 
{\it NMI}~\cite{StrehlG02}, 
{\it Shen-F}~\cite{shen-F},
{\it F1}~\cite{Crestani01logicand}.



\textbf{Adjusted Rand Index (ARI)} is commonly used in cluster analysis to measure the degree of agreement between two data clusters. An index value that is closer to 0 represents random labeling, where a value of 1 indicates that the data clusters are identical. 
\textbf{Normalized Mutual information (NMI)} is a normalization of the Mutual Information (MI) score to scale the results between 0 (no mutual information) and 1 (perfect correlation).
\textbf{NMI} evaluates the distribution consistency of predicted and true clustering, regardless of the sorting of the clustering.  
\textbf{F1} calculates the number of perfectly matched dialogs and is considered as the most strict measure. 
\textbf{Shen-F} defines a weighted sum of F1 scores over all dialogs. The F1 weights are calculated based on the number of utterances in the dialogs. 

\begin{table}[t]

\centering
\begin{threeparttable}
 \resizebox{0.95\columnwidth}{!}{
\begin{tabular}{c|c|c|c|c|c|c|c|c|c|c}
\toprule
& P & R & \textbf{F1} & 1-to-1 &$loc_3$ & MAP & MRR & \textbf{NMI} & \textbf{ARI} & \textbf{Shen-F}\\
\hline
Weighted-SP& & & \Checkmark & & & &&&&\Checkmark\\
\hline 
ME Classifier&\Checkmark&\Checkmark& && & &&&&\\
\hline 
CISIR & & & & \Checkmark & \Checkmark & &&&&\Checkmark\\
\hline 
BiLSTM & & & \Checkmark & & & \Checkmark &&\Checkmark&\Checkmark&\Checkmark\\
\hline 
FF & \Checkmark & \Checkmark & \Checkmark & & \Checkmark &&\Checkmark&&&\Checkmark\\
\hline
BERT &  & & \Checkmark & & &&&\Checkmark&\Checkmark&\Checkmark \\
\hline 
E2E & & & \Checkmark & & & &&\Checkmark&\Checkmark&\Checkmark \\
\hline 
PtrNet & \Checkmark &\Checkmark &\Checkmark & & & &&\Checkmark&\Checkmark& \\
\bottomrule
\end{tabular}}
\end{threeparttable}
\caption{Selection of representative measures.}
\label{metrics_selection}
\end{table}


\subsection{Experiments}

We conduct two experiments to evaluate how effective SOTA models are at disentangling software-related chat. 

\begin{figure}[t]
\centering
\includegraphics[width=\columnwidth]{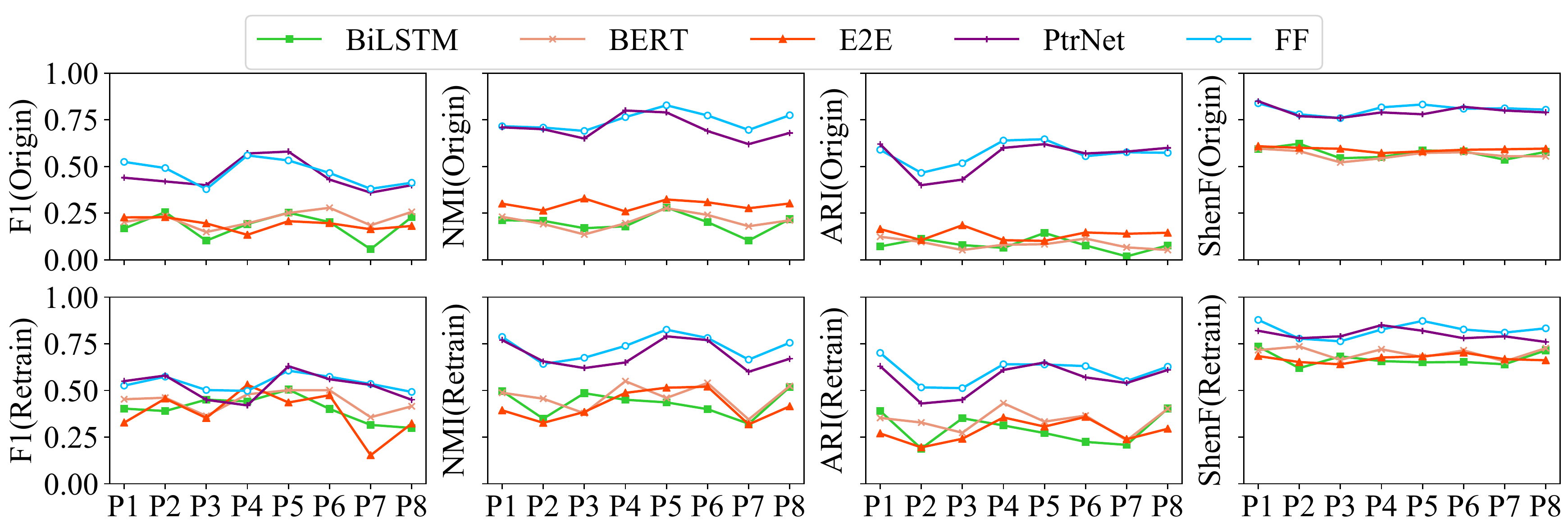}
\caption{Performance comparison of five SOTA models.} 
\label{RQ1_Model_Chosen}
\end{figure}

\textbf{Experiment \#1 (Original).} 
{We use the five original SOTA models as described and trained in the existing literature to disentangle software-related chat.}
As shown in Figure \ref{RQ1_Model_Chosen}, 
\textit{BiLSTM}, \textit{Bert}, and \textit{E2E} models can only achieve medium-level scores on Shen-F, with low performances on F1, NMI and ARI.
\textit{FF} and \textit{PtrNet} models significantly\footnote{All the $p$ values in significance T-test are $p<0.05$.} outperform the other three models. However, further analysis shows no significant difference ($p=0.56$) between \textit{FF} and \textit{PtrNet} models. Since \textit{FF} model achieves slightly higher performances than \textit{PtrNet} model on average, we consider \textit{FF} model performs the best at disentangling software-related dialogs in this experiment. Specifically, \textit{FF} model can achieve high scores on clustering measures (avg(NMI)=0.74 and avg(Shen-F)=0.81), medium-level scores on perfectly matching (avg(F1)=0.47), and pairwise clustering similarity (avg(ARI)= 0.57).


\textbf{Experiment \#2 (Retraining).}
We then retrain the five SOTA models on software-related chat. 
{We use ``retrain" to refer to the fine-tuning as well as the training on the new data. Specifically, the original Bert, E2E, PtrNet, and BiLSTM utilize pre-training for contextual embedding, so we fine-tune the four SOTA models on the new data. Since FF is a feedforward network, we train it on the new data directly.}
For each SOTA model, we retrain with seven projects and evaluate with the eighth project. 
In order to study the significance of the performance improvement contributed by retraining, we conduct five paired t-tests between each original model and its retrained model. The $p$ values indicate that the retrained \textit{FF} and \textit{PtrNet} do not have significant improvement ($p_1=0.43$, $p_2=0.62$) to the original model. 
More specifically, the retraining strategy does not guarantee performance improvement from their original models. As shown in Figure \ref{RQ1_Model_Chosen}, while the retrained FF model increases its F1, ARI, and Shen-F by 0.03, 0.03, and 0.01 on average respectively, the average NMI score decreases by 0.01. The retrained \textit{PtrNet} model displays a similar trend. The F1 and ARI increase by 0.07 and 0.01 respectively on average, while the average Shen-F shows no change and the average NMI score decreases by 0.02.
On the contrary, the retrained \textit{BiLSTM}, \textit{BERT} and \textit{E2E} models significantly ($p_3=10^{-5}$, $p_4=10^{-6}$, $p_5=10^{-3}$) outperform the original ones. 
As shown in Figure \ref{RQ1_Model_Chosen}, the retrained \textit{BERT} model obtains the largest performance improvement among the three models: the average F1, NMI, ARI and Shen-F increases by 0.22, 0.25, 0.26, and 0.13 respectively.
However, the retrained \textit{FF} and \textit{PtrNet} models still significantly outperform the other three retrained models.
%

\textbf{Finding:} 
Both \textit{FF} and \textit{PtrNet} models outperform the rest in both experiments. However, we observe that the retraining strategy is unable to contribute significantly to the original models in terms of performance improvement. 
Since SOTA models are built with general dialogs, retraining still cannot effectively address the challenges that SOTA models face in disentangling SE dialogs, such as the lack of understanding of domain-specific terminology and the inefficiency in disentangling complex dialogs with mixing up topics.
Moreover, after combining dropout with early stopping to avoid risk of overfitting, we observed that convergences were already achieved at EPOCH 10-15, thus we believe that the retraining performance could not be better even given more data.

In this study, we recommend adopting the original \textit{FF} model as the best model to disentangle software-related dialogs, as the original \textit{FF} model can achieve slightly higher performances on average than the original \textit{PtrNet} model. Despite the original \textit{FF} model being the best, the average F1 score is only 0.47, which indicates a relatively low perfectly matching score.

 \section{Human Satisfaction Measures}
 

Since the goal of dialog disentanglement is to make it easy for users to find information from entangling dialogs, it is essential to evaluate the quality of the disentangled results against human satisfaction. Various measures (as introduced in Section 3.2) have been proposed to calculate how close the disentangled results are from the ground-truth data. To understand how well these measures reflect human satisfaction, we conduct an experiment to compare these existing measures with manually-scored human satisfaction.



\begin{figure*}[t]
\centering

\begin{minipage}[t]{0.39\textwidth}
\centering
\includegraphics[width=0.9\textwidth]{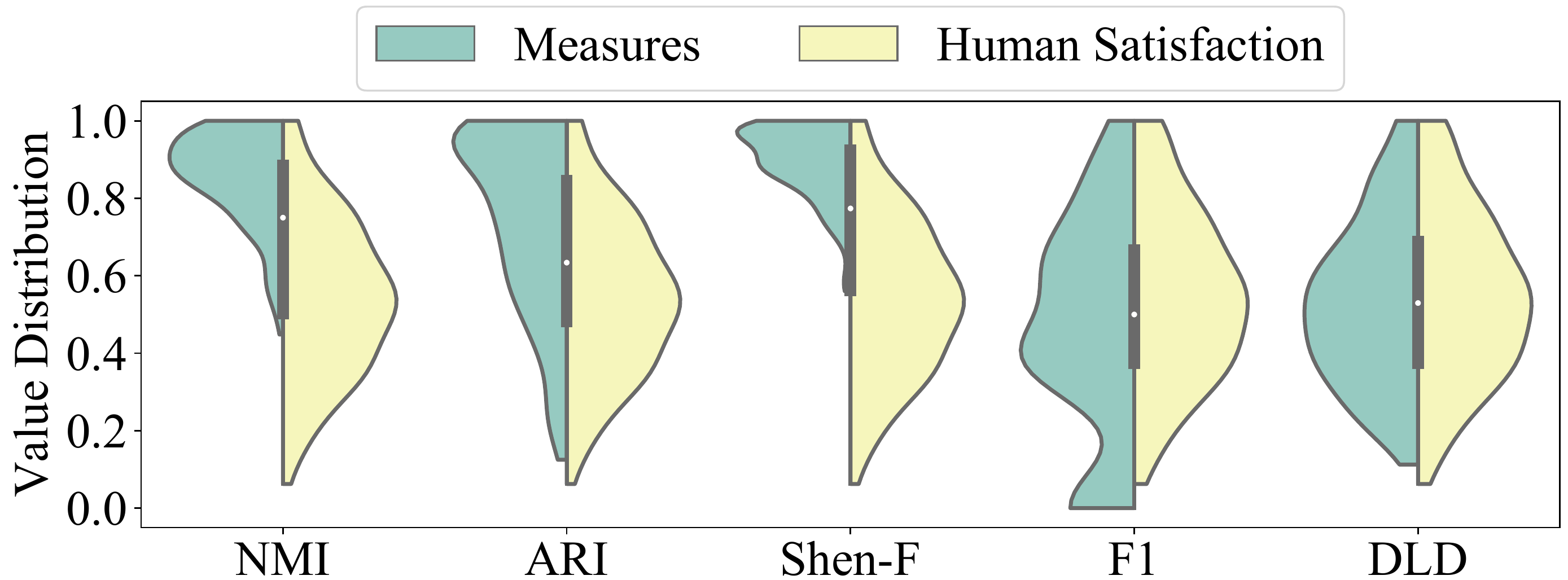}
\caption{Value distribution of the five different measures and human satisfactions.}
\label{all_metrics}
\end{minipage}
\begin{minipage}[t]{0.39\textwidth}
\centering
\includegraphics[width=0.95\textwidth]{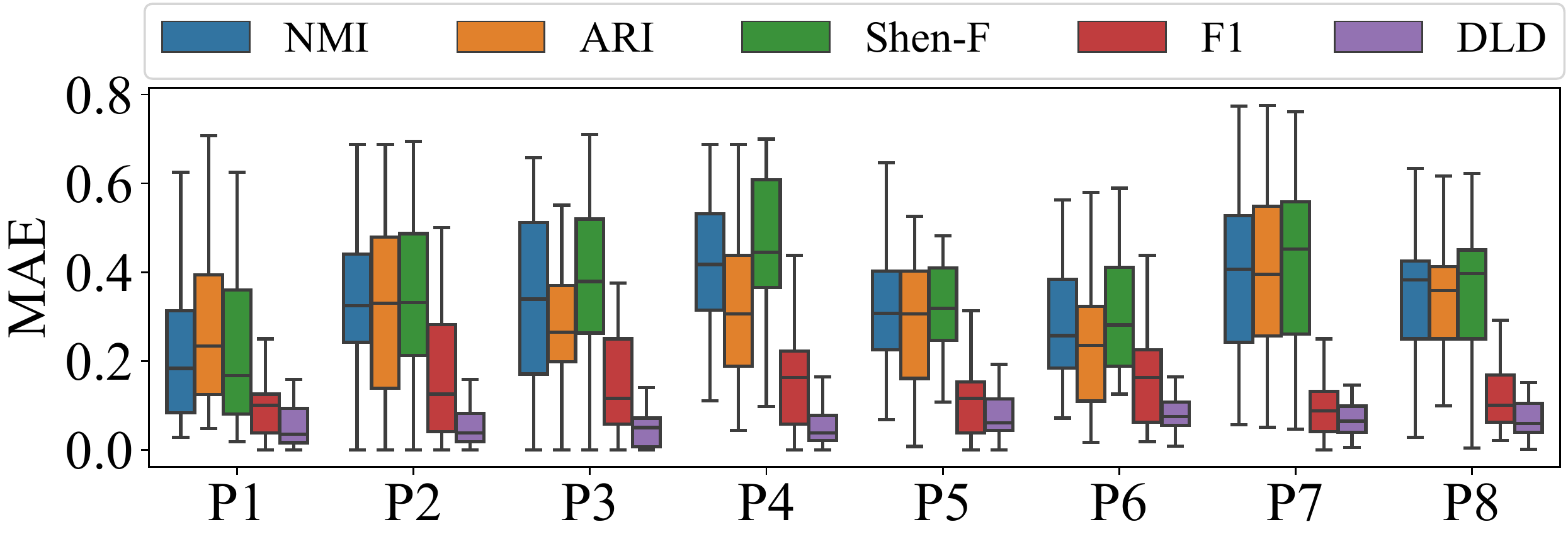}
\caption{MAE comparison within projects.}
\label{project_metrics}
\end{minipage}
\begin{minipage}[t]{0.2\textwidth}
\centering
\includegraphics[width=0.92\textwidth]{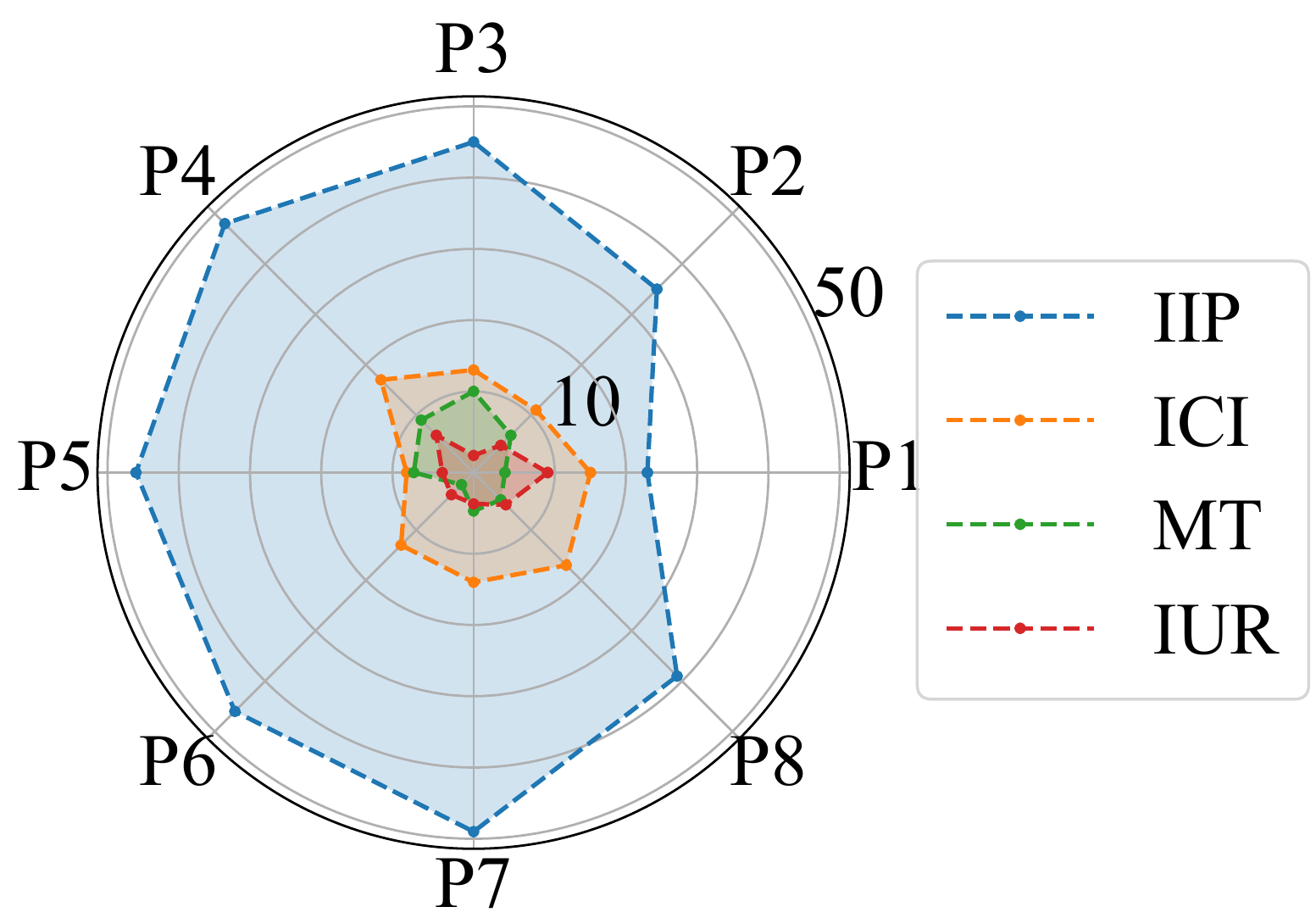}
\caption{Distributions of bad cases in projects.}
\label{bad_cases}
\end{minipage}


\label{metrics_gt}
\end{figure*}


\subsection{Measures Analysis}

\subsubsection{Manually Scoring Human Satisfaction}
\begin{table}[t]
\scriptsize
\centering
\begin{threeparttable}
 \resizebox{0.95\columnwidth}{!}{
\begin{tabular}{c|cc|c|ccc}
\toprule
Category & \multicolumn{2}{c|}{Error$^1$} & Correlation$^2$& \multicolumn{3}{c}{Hypothesis$^3$} \\
\hline
Analysis & RMSE & MAE & PEA & IST & PST & ANOVA \\
\hline
NMI &0.38 &0.34 & 0.08 & $10^{-55}$ & $10^{-46}$ & $10^{-55}$  \\
ARI & 0.37 & 0.32 & 0.02 & $10^{-19}$ & $10^{-17}$ & $10^{-19}$ \\
Shen-F & 0.41 & 0.36 & 0.17 & $10^{-69}$ & $10^{-59}$ & $10^{-69}$\\
F1 & 0.19 & 0.14 & 0.85 & $10^{-4}$ & $10^{-14}$ & $10^{-4}$ \\
\hline
$DLD$ & {\bf 0.08} & {\bf 0.07} & {\bf 0.92} & 0.51 & 0.31 & 0.51\\
\bottomrule
\end{tabular}}
 \begin{tablenotes}
        \item $^1$ Negative correlation with effectiveness. $^2$ Positive correlation with effectiveness. $^3$ Significance test, where $p<0.05$ indicates significant difference.
      \end{tablenotes}
\end{threeparttable}
\caption{Deviation analysis results between existing measures and human satisfaction scores.} 
\label{metrics_eval}
\end{table}

We build two teams to manually score human satisfaction of the disentangled dialogs produced by the original FF model. Each team consists of one Ph.D. candidate and one developer. All of them are fluent English speakers and have done either intensive research work with software development or have been actively contributing to open-source projects. We leverage a five-point Likert scale~\cite{likert} to score human satisfaction. The rating scale varies from ``strongly unsatisfied" to ``strongly satisfied", with the values of 1 to 5 respectively. Each disentangled dialog receives two scores from two team members, and we host discussions for dialogs with inconsistent ratings. {The average Cohen’s Kappa score \cite{cohen_kappa} of our study is 0.82, indicating that the participants highly agree with each other.}



\subsubsection{Deviation Analysis}
We then conduct a quantitative analysis to compare the differences between the existing measures and the manually-scored human satisfaction. 

Since F1, ARI, NMI, and Shen-F all range from 0 to 1, while manually-scored human satisfaction scales from 1 to 5, to compare the differences, we normalize human satisfaction into $score\in[0,1]$ with a linear function: $score=0.25\times(score-1)$. We then analyze the deviation between existing measures and manually-scored human satisfaction in following aspects: 

(1) Error analysis: We calculate {\it RMSE} (Root-mean-square Error) and {\it MAE} (Mean Absolute Error)~\cite{mae_rmse} as the error deviation between two samples.

(2) Correlation analysis: We calculate {\it Pearson} (PEA) correlation coefficient between two samples.

(3) Hypothesis testing: {We perform {\it Independent Sample T-Test} (IST), {\it Paired Sample T-Test} (PST), and {\it Analysis of Variance} (ANOVA) to measure and validate the deviation between two samples:}

$\bullet$ {\it Independent Sample T-Test} (IST): $H_0: \mu_1=\mu_2, H_1: \mu_1\neq\mu_2$.

$\bullet$ {\it Paired Sample T-Test} (PST): $H_0: \mu_Z=0, H_1: \mu_Z\neq0$, where $Z=X_1-X_2$.

$\bullet$ {\it Analysis of Variance} (ANOVA): $H_0: \sigma_1^2=\sigma_2^2, H_1: \sigma_1\neq\sigma_2$

Table \ref{metrics_eval} shows the deviation analysis results. By comparing the error and correlation results of NMI, ARI, Shen-F and F1, we find that F1 is the most similar to human satisfaction scores, with the lowest Error ($RMSE=0.19$, $MAE=0.14$) and the highest correlation ($PEA=0.85$). However, F1 is not statistically acceptable since the hypothesis testing shows a significant difference with human satisfaction ($p<0.05$, \textit{reject}). Figure \ref{all_metrics} visualizes the value distribution of measures and human satisfaction. We can see that NMI, ARI, and Shen-F are likely to overrate the disentanglement results, while F1 slightly underrates them.  

\textbf{Finding:} We can conclude that existing measures are unable to reflect human satisfaction accurately, thus a new measure is needed.

\subsection{A Novel Measure: DLD}


We leverage {\it Levenshtein Distance}~\cite{levendist} and {\it Levenshtein Ratio}~\cite{levendist2} to estimate the editing effort of dialog-to-dialog transition (i.e.: {\it Delete}, {\it Insert} and {\it Update}). 
Given the ground-truth disentanglement dialogs $D_T$ and the predicted disentanglement dialogs $D_P$, {we first calculate the differences between $D_P$ and $D_T$.}
\begin{equation}
    \Delta(D_T,D_P)=|D_T-D_P|+|D_P-D_T|.
\end{equation}
where $|D_i-D_j|$ denotes the number of utterances in $D_i$ that are not included in $D_j$.
We then perform the Sigmod function $\sigma$ to {normalize} the absolute value {into $(0,1)$}.
\begin{equation}
\sigma(\Delta(D_T,D_P), \eta) = 1/(1+e^{\Delta(D_T,D_P)-\eta}),
\end{equation}
{where $\eta$ is the threshold number of utterances that indicates good disentanglement or not.}
{We define {\it Dialog Levenshtein Revision} as follows:}
\begin{equation}
    DLR_v=\mathbb{E}[\sigma(\Delta(D_T,D_P), \eta)],
\end{equation}
{where $\mathbb{E}$ denotes the expectation of a collection of  dialog pairs $<D_P, D_T>$.} 
Since $DLR_v$ measures the size of absolute revisions, we further define {\it Dialog Levenshtein Ratio} to measure the proportion of revisions in one dialog: 
\begin{equation}
DLR_t=\mathbb{E}[1-\Delta(D_T,D_P)/(|D_T|+|D_P|)].
\end{equation}

By taking both $DLR_v$ and $DLR_t$ into consideration, we define {\it Dialog Levenshtein Distance} as a novel measure for human satisfaction:
\begin{equation}
    DLD=\lambda DLR_t+(1-\lambda)DLR_v, 0\leq\lambda\leq 1.
\end{equation}


\subsubsection{Effectiveness of DLD}
{By tuning the values of $\eta$ and $\lambda$, we observe that \textit{DLD} can achieve the smallest deviation when $\eta=5$ and $\lambda=0.8$.}
The last row of Table \ref{metrics_eval} shows the error, correlation, and hypothesis testing analysis results between \textit{DLD} and manually-scored human satisfaction. Compared with other measures, \textit{DLD} achieves the lowest error ($RMSE=0.08$, $MAE=0.07$) and highest correlation ($PEA=0.92$). Moreover, figure \ref{project_metrics} illustrates that \textit{DLD} can achieve the lowest error across all projects.
The $p$ values of the three hypothesis tests are all over 0.05, which indicates that \textit{DLD} shows no significant differences when compared to manually-scored human satisfaction. Figure \ref{all_metrics} visualizes such value distribution. The \textit{DLD} shows the most symmetric shape, which indicates that our measure is the most accurate in matching the manually-scored human satisfaction.  


\textbf{Finding:} {While the existing measures evaluate the similarity between the predicted and true disentanglement results based on perfectly matching or clustering distribution, \textit{DLD} incorporates the Levenshtein distance to quantitatively measures such similarity. When compared to the existing measures, \textit{DLD} can more accurately measure human satisfaction.}
{We consider the reason is that, edit distance could
reflect the completeness of information in relation to the dialog
structure whereas the existing measures do not.}
{Moreover, the \textit{DLD} can also be extensively used for general dialog disentanglement since it dose not involve any SE-specific knowledge. }

\section{Bad Cases Analysis}



We perform an in-depth analysis on the ``bad case'' disentangled dialogs that received human satisfaction $score\leq3$ to further investigate why the disentangled dialogs are unsatisfying. 
We classify these disentangled dialogs by using open card sorting~\cite{cardsorting1}. Similar to the process of scoring human satisfaction, we group participants into two teams to cross-inspect the classifications of bad cases. Discussions are hosted for inconsistent classifications.


As the result, we identify four categories: {\it Ignoring User Relationships, Mixing up Topics, Ignoring Contextual Information,} and {\it Ignoring Interaction Patterns}. As shown in Figure \ref{bad_cases}, all four bad cases occur in every sampled project. The most commonly occurred bad case (64\%) is {\it Ignoring Interaction Patterns}. 
{These bad cases might also occur in general dialogs with different distributions.}
To better understand each bad case, we further elaborate definitions with examples.



\subsection{Ignoring Interaction Patterns (IIP: 64\%)}

\textbf{Definition}: The utterances in disentangled dialogs are incorrect due to missing the utterances that comply with the interaction patterns.

\noindent
\textbf{Example 1.}

\begin{figure}[H]
\centerline{\includegraphics[width=0.9\columnwidth]{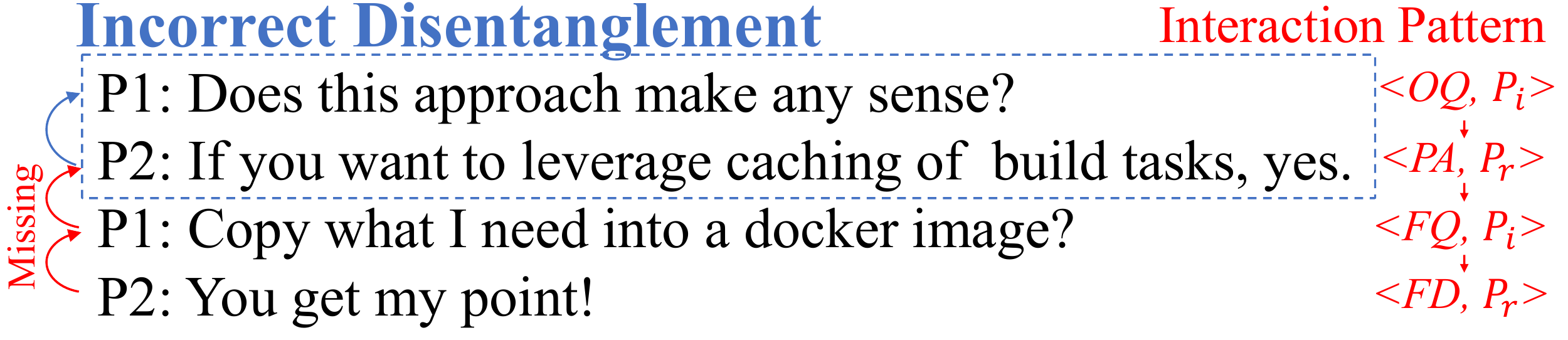}}
\label{eg-f}
\end{figure}


Three interaction patterns have been observed when analyzing the content of these unsatisfying disentangled dialogs. We leverage the user intents proposed by Bayer {\it et al.}~\shortcite{ese20dialogue} and Qu {\it et al.}~\shortcite{userintent19}. These user intents represent the utterance intention, which indicates what the participants want to convey when posting messages. Each pattern is a sequence of $<$Intent, Role$>$, where Intent=\{OQ (Origin Question), PA (Potential Answer), FQ (Follow-up Question), FD (Further Detail), CQ (Clarify Question)\}, and Role=\{$P_i$(Dialog Initiator), $P_r$ (Respondent)\}. The interaction patterns can be described as follows:

(1) \textit{Direct Answer} ($<$OQ, $P_i>$, $<$PA, $P_r>$):
$P_i$ initiates the dialog with OQ,  $P_r$ then directly answers with $PA$.

    
(2) \textit{Clarifying Answer} ($<$OQ, $P_i>$, $<$PA, $P_r>$, $<$FQ, $P_i>$, $<$FD, $P_r>$): 
$P_i$ initiates the dialog with OQ, a potential answer PA is posted by $P_r$. Then a sequence of follow-up questions is posted by dialog initiator $P_i$ to clarify until the answer is fully understood and accepted by $P_i$.
    
(3) \textit{Clarifying Question} ($<$OQ, $P_i>$, $<$CQ, $P_r>$, $<$FD, $P_i>$, $<$PA, $P_r>$):
$P_i$ initiates the dialog with OQ, a set of clarifying questions CQ is posted by $P_r$ to clarify the original question until the original question is fully understood by $P_r$. Then a potential answer $PA$ is posted by $P_r$.

The dialog in Example 1 illustrates the \textit{Clarifying Answer} pattern. After $P_2$ provides a potential answer, $P_1$ asks a follow-up question to clarify the answer. Then $P_2$ answers the follow-up question. In this case, the existing disentanglement models fail to predict that the four utterances should be in the same dialog. Instead, only the first two utterances are considered as one dialog. 
Thus, we recommend that disentanglement models can be optimized by identifying utterances that display the above interaction patterns.




\subsection{Ignoring Contextual Information (ICI: 21\%)}
\textbf{Definition}: The utterances in disentangled dialogs are incomplete due to the missing of contextual-related utterances.

\noindent
\textbf{Example 2.}

\begin{figure}[H]
\centerline{\includegraphics[width=0.9\columnwidth]{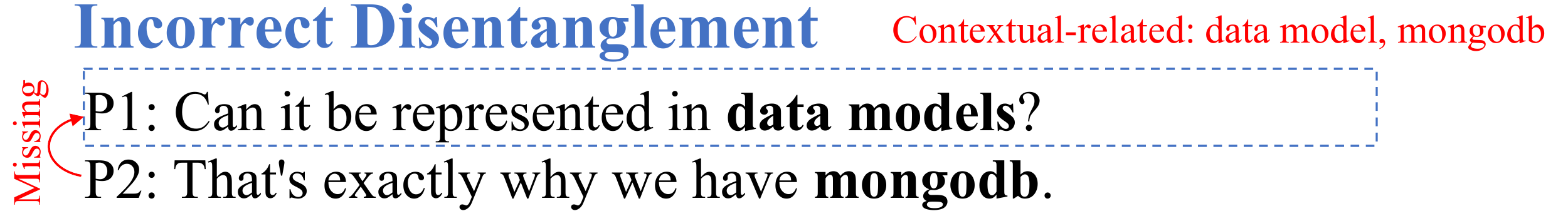}}
\label{eg-f}
\end{figure}


Example 2 shows an incorrect disentanglement that misses respondent utterance. $P_1$ asks about the usage of data models and $P_2$ replies with MongoDB with no further clarifications. Although we know that MongoDB is a data model and these two utterances are highly related, since the existing disentanglement models often use textual similarity, these two utterances appear to be as less related. Thus, lacking such contextual information will often lead to incorrect disentanglement. Therefore, we recommend disentanglement models to incorporate contextual similarity, such as pre-trained word vectors GloVe~\cite{glove,lowedataset,basic_corpus}.

\subsection{Mixing up Topics (MT: 9\%)} 
\textbf{Definition}: The disentangled dialogs contain multiple topics that are discussed by the same group of participants.

\noindent
\textbf{Example 3.}

\begin{figure}[H]
\centerline{\includegraphics[width=0.9\columnwidth]{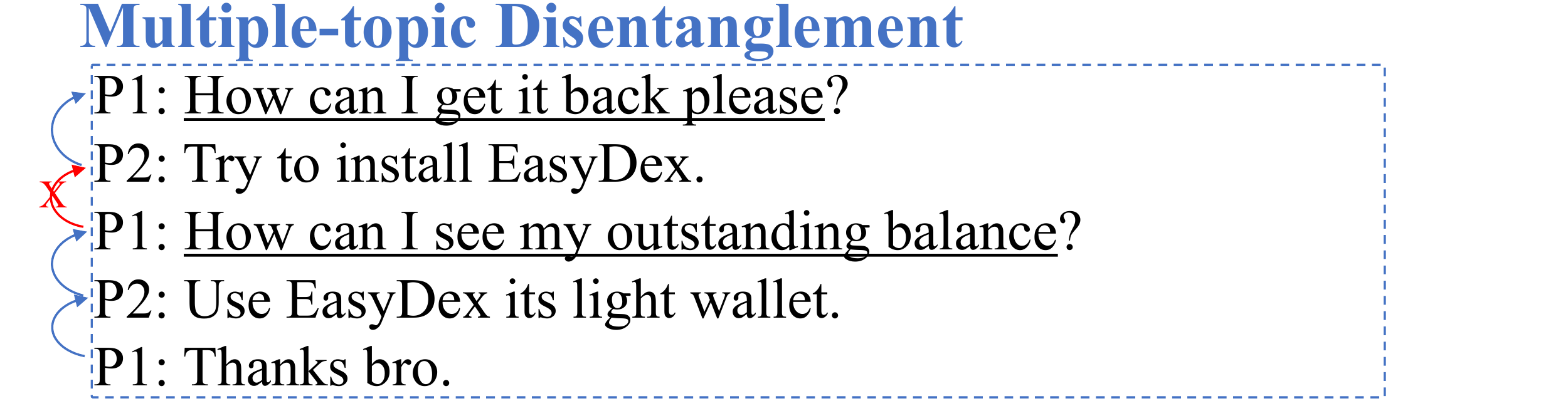}}
\label{eg-f}
\end{figure}


Example 3 shows an incorrect disentangled dialog with mixed up topics in one conversation. $P_1$ asks two questions that belong to different topics while $P_2$ provides separate answers to each question. The first two utterances talk about the installation of EasyDex, while the following two utterances talk about the usage of EasyDex. Such disentangled dialog violates the single-topic principle of dialog disentanglement~\cite{dialogue_principle}. Therefore, we recommend integrating topic extraction algorithms, such as {\it LDA}~\cite{lda} into disentanglement models to separate topics.

\subsection{Ignoring User Relationships (IUR: 6\%)}
\textbf{Definition}: The utterances in disentangled dialogs are incorrect due to the lack of understanding on the relationships among the participants.

\noindent
\textbf{Example 4.}

\begin{figure}[H]
\centerline{\includegraphics[width=0.9\columnwidth]{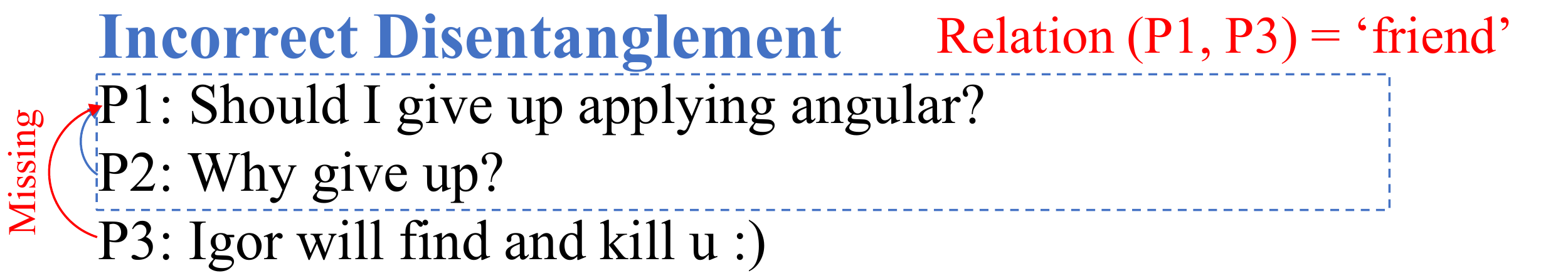}}
\label{eg-t}
\end{figure}

Example 4 shows an incorrect disentanglement when the relationship between participants is ignored. There are three utterances in this dialog: $P_1$ posts a question, $P_2$ and $P_3$ reply with two distinct answers. The existing disentanglement models only predict $P_1$ and $P_2$ as one dialog, while excluding $P_3$. Without understanding the relationship between $P_1$ and $P_3$, the answer posted by $P_3$ seems to be irrelevant. However, by analyzing their post histories, we find that 76\% of the utterances posted by $P_1$ are associated with the utterances posted $P_3$. With such frequent interaction, we consider that $P_1$ and $P_3$ have a close relationship. Thus, $P_3$ is very likely to be in the same dialog as $P_1$ and $P_2$. 
Therefore, we recommend that disentanglement models can use the participants' relationship and collaboration history to improve performance.





\section{Conclusion}

{In this paper, we evaluate five state-of-the-art dialog disentanglement models on software-related dialogs to investigate how these models can be used in the context of software engineering. To acquire the best performing model, we conduct two experiments with the original and the retrained models respectively. Considering the trade-offs between training effort and performances, the results show that the original FF model is the best model on disentangling software-related dialogs. Although the original FF model has the best performance, the evaluation measures do not accurately reflect human satisfaction. Thus, we introduce a novel measure \textit{DLD}. Compared to other measures, \textit{DLD} can more accurately measure human satisfaction. Finally, we investigate the reasons why some disentangled dialogs are unsatisfying. By classifying these disentangled dialogs, we identify four common bad cases. 
We believe that our study can provide clear directions on how to optimize existing disentanglement models.

}

\section*{Acknowledgments}


This work is supported by the National Key Research and Development Program of China under Grant No. 2018YFB1403400, the National Science Foundation of China under Grant No. 61802374, 62002348, and 62002348, and Youth Innovation Promotion Association CAS.
\bibliographystyle{named}
\bibliography{ijcai21}
\end{document}